\documentclass[12pt,preprint]{aastex}
\usepackage{amsmath}
\begin{document}
\title{Polarization of Astrophysical Events with Precessing Jets}
\author{Mi-Xiang Lan$^{1,2}$, Rui Xue$^{2}$, Dingrong Xiong$^{3}$, Wei-Hua Lei$^{4}$, Xue-Feng Wu$^{1}$, and Zi-Gao Dai$^{2,5}$}
\affil{$^{1}$Purple Mountain Observatory, Chinese Academy of Sciences,
Nanjing 210008, China; lanmixiang@126.com, xfwu@pmo.ac.cn \\
$^{2}$School of Astronomy and Space Science, Nanjing University, Nanjing 210093, China; dzg@nju.edu.cn\\
$^{3}$Yunnan Observatories, Chinese Academy of Sciences, Kunming 650216, China\\
$^{4}$School of Physics, Huazhong University of Science and Technology, Wuhan 430074, China\\
$^{5}$Key Laboratory of Modern Astronomy and Astrophysics (Nanjing
University), Ministry of Education, China \\}

\begin{abstract}
A central compact object (CCO, e.g. a black hole) with an accretion disk has been suggested as the common central engine of various astrophysical phenomena, such as gamma-ray bursts (GRBs), tidal disruption events (TDEs) and  active galactic nuclei (AGNs). A jet powered by such a system might precess due to the misalignment of the angular momenta of the CCO and accretion disk. Some quasi-periodic behaviors observed in the light curves of these phenomena can be well interpreted within the framework of a precessing jet model. In this paper, we study the emission polarization of precessing jets in the three kinds of phenomena. The polarization angle also shows a gradual change for the synchrotron emission in both the random and toroidal magnetic field configurations with the precessing jet, while it can only change abruptly by $90^\circ$ for the non-precessing top-hat jet. Polarization properties are periodic due to the assumptions made in our model. The polarization observations are crucial to confirm the precession nature of jets in GRBs, TDEs and AGNs.

\end{abstract}

\keywords{gamma-ray burst: general --- galaxies: jets --- magnetic fields --- polarization --- radiation mechanisms: nonthermal}

\section{Introduction}
Relativistic Jets are involved in many astrophysical phenomena, such as gamma-ray bursts (GRBs; Lipunov et al. 2001; Frail et al. 2001; Rossi et al. 2002; Zhang \& M\'esz\'aros 2002), tidal disruption events (TDEs; Bloom et al. 2011b; Burrows et al. 2011; Levan et al. 2011; Zauderer et al. 2011; Cenko et al. 2012; Brown et al. 2015; Pasham et al. 2015) and active galactic nuclei (AGNs; Krolik 1999; Bottcher et al. 2012; Beckmann \& Shrader 2012). Central engines of these events might be central compact objects (CCOs, e.g. black holes) plus accretion disk systems (Pringle 1981; Rees 1988; Abramowicz et al. 1989; Narayan et al. 1992; Narayan \& Yi 1994; Yuan 2001; Sadowski et al. 2014; Jiang et al. 2014; Yuan \& Narayan 2014). Since the angular momentum of the accretion disk is not necessarily parallel to that of the CCO, a jet powered by such a central engine may precess (Sarazin et al. 1980; Lu 1990; Lu \& Zhou 2005).

For GRBs, a precessing jet model has been proposed to interpret the variability of prompt emissions (Portegies Zwart et al. 1999; Reynoso et al. 2006; Lei et al. 2007), the spectral evolution during prompt phase (Liu et al. 2010), and the quasi-periodic behavior in some GRB afterglows (e.g., GRB 060206 and GRB 130925A; Liu et al. 2008; Hou et al. 2014). For a black hole-neutron star (BH-NS) merger, a typical period of $\sim30-100\,\rm ms$ is predicated by this model  (Stone et al. 2013). Therefore, it is reliable to distinguish progenitor models (i.e, BH-NS or NS-NS mergers) by inspecting the temporal behavior of short GRBs.

TDEs, i.e., disruption of a star or sub-stellar object by tidal forces when it passes close enough by a super-massive black hole (SMBH), provide an effective way to discover quiescent SMBHs. In events, about half of the stellar material will be accreted onto the SMBH, producing a flare of electromagnetic radiation (Rees 1988). Swift J16449.3+573451 (``Sw J1644+57" hereafter) is the first discovered TDE with a relativistic jet towards the earth. The X-ray light curve after the peak is consistent with the $t^{-5/3}$ decay law, as expected by the standard TDE picture (Bloom et al. 2011a, 2011b; Burrows et al. 2011; Cannizzo, Troja \& Lodato 2011; Barres de Almeida \& De Angelis 2011; Socrates 2012). Two kinds of quasi-periodic oscillations (QPOs) have been reported for this event: a $\sim$200 s QPO (Reis et al. 2012) and a 2.7 day QPO (Burrows et al. 2011). Many models have been proposed to interpret such observational features. For example, the disk resonance mechanism (Abramowicz \& Liu 2012) and the jet-magnetically-arrested-disk QPO mechanism (Tchekhovskoy et al. 2014) have been applied to explain the 200 s QPO. A jet-precessing model was proposed by Lei et al. (2013) to interpret the 2.7 day QPO. Later, Wang et al. (2014) developed a precessing two-component jet model to interpret the two QPOs (200s and 2.7 day) in Sw J1644+57.

AGNs are the most luminous persistent sources in the universe. The year-like quasi-periodicity in three BL Lac objects was analyzed recently by Sandrinelli et al. (2018). There are two popular models for the quasi-periodicity, i.e., the binary supermassive black hole system (BSMBH; Lehto \& Valtonen 1996; Graham et al. 2015) and the precession process (Camenzind \& Krockenberger 1992; Marscher 2014; Raiteri et al. 2017). However, the BSMBH model was unfavored by the gravitational wave background observations (Wang et al. 2015).

Polarization evolution is particularly important because it may provide useful information about the jets and central engines. In our previous work (Lan, Wu \& Dai 2016), we calculated the polarization evolution of a non-precessing GRB jet during the early afterglow phase. We found that polarization angle (PA) can change gradually for the aligned magnetic field configuration (MFC) which is not expected for the toroidal configuration. Our result thus provided a probe for the central engine of GRBs since a magnetar central engine tends to produce an aligned field configuration (Spruit et al. 2001) while a BH central engine tends to launch a toroidal field dominated jet via the Blandford-Znajek mechanism (hereafter BZ, Blandford \& Znajek 1977). In this paper, we extend our study on polarization evolution to a precessing jet and to three different kinds of events (i.e., GRBs, TDEs and AGNs). In general, we consider three polarization models, synchrotron emission in an ordered aligned magnetic field (SOA), synchrotron emission in an ordered toroidal magnetic field (SOT) and synchrotron emission in a random magnetic field (SR).

This paper is arranged as follows. In Section 2, the models are described. In Section 3, we present our numerical results for the polarization evolution of precessing jets in GRBs, TDEs and AGNs. We summarize the results in Section 4 with some discussions.

\section{The Models}

\subsection{GRB and TDE Cases}
We assume that the bulk Lorentz factor ($\gamma$) of a relativistic jet does not vary during one period of precession. The GRB spectrum is described with the Band function (Band et al. 1993). For the spectrum of TDE, we also assume the Band function as indicated by the observations of Sw J1644+57 (Burrows et al. 2011). The synchrotron emission is assumed here. The spectrum function $g(\tilde{x})$, the local polarization degree (PD, $\pi_0$) in an ordered magnetic field and the spectral index ($\alpha_1$) are given by (Toma et al. 2009),
\begin{equation}
g(\tilde{x})=\begin{cases}
\tilde{x}^{-\alpha_s}e^{-\tilde{x}}, & \text{$\tilde{x}<\beta_s-\alpha_s$}, \\ \tilde{x}^{-\beta_s}(\beta_s-\alpha_s)^{\beta_s-\alpha_s}e^{\alpha_s-\beta_s}, & \text{$\tilde{x}\geq\beta_s-\alpha_s$},
\end{cases}
\end{equation}
\begin{equation}
\pi_0=\begin{cases}
(\alpha_s+1)/(\alpha_s+5/3), & \text{$\tilde{x}<\beta_s-\alpha_s$}, \\ (\beta_s+1)/(\beta_s+5/3), & \text{$\tilde{x}\geq\beta_s-\alpha_s$},
\end{cases}
\end{equation}
\begin{equation}
\alpha_1=\begin{cases}
\alpha_s, & \text{$\tilde{x}<\beta_s-\alpha_s$}, \\ \beta_s, & \text{$\tilde{x}\geq\beta_s-\alpha_s$},
\end{cases}
\end{equation}
where $\tilde{x}\equiv\nu'/\nu'_0$. The quantities in the comoving frame of the jet are denoted with a prime. $E'_p=h\nu'_0$ is the break energy and $\nu'=\nu(1+z)\gamma(1-\beta\cos\theta)$. $\nu$ is the observational frequency, $z$ is the redshift of the source, $\beta$ is the dimensionless speed (in unit of $c$, the speed of light) of the jet, $\theta$ is the angle between the velocity of a small emitting region in the jet\footnote{The lateral expansion is not considered and thus the velocity is along the radial direction.} and the line of sight (LOS), $\alpha_s$ and $\beta_s$ are the low-energy and high-energy spectrum index, respectively.

For the SR model, we have (Toma et al. 2009)
\begin{equation}
F_{\nu}=\frac{1+z}{d_L^2}\frac{R^2A_0}{T_0}\int_0^{\theta_j+\theta_V}d\theta \mathcal{D}^2f(\nu')\sin\theta \langle (\sin\theta'_B)^{\alpha_1+1}\rangle2\Delta\phi ,
\end{equation}
and
\begin{equation}
Q_{\nu}=\frac{1+z}{d_L^2}\frac{R^2A_0}{T_0}\int_0^{\theta_j+\theta_V} d\theta\mathcal{D}^2f(\nu')\sin\theta\langle (\sin\theta'_B)^{\alpha_1+1}\rangle\pi_p\int_{-\Delta\phi}^{\Delta\phi}d\phi \cos(2\chi_p),
\end{equation}
where $F_\nu$ is the time-averaged flux and $Q_\nu$ is the Stokes parameter. It can be proved that the Stokes parameter $U_{\nu}=0$ for a random magnetic field confined in the shock plane (Toma et al. 2009; Lan, Wu \& Dai 2016). $d_L$ is the luminosity distance of the source, $R$ is the radius of the emission region, the normalization $A_0$ is a constant with units of erg cm$^{-2}$ str$^{-1}$ Hz$^{-1}$, $T_0$ is the duration of a GRB, $\theta_j$ is the half-opening angle of the jet, $\theta_V$ is the observational angle (i.e., the angle between the jet axis and LOS), $\mathcal{D}=1/\gamma(1-\beta\cos\theta)$ is the Doppler factor. We denote $\bar{q}=-\pi_0\langle (\sin\theta'_B)^{\alpha_1+1}\cos(2\phi'_B)\rangle$.
The local PD can be expressed as $\pi_p=|\bar{q}|/\langle (\sin\theta'_B)^{\alpha_1+1}\rangle$. $\theta'_B$ and $\phi'_B$ are the polar and azimuthal angle of the magnetic field in the right-handed orthogonal coordinate system $\hat{1}\hat{2}\hat{k'}$ with $\hat{k'}$ the comoving LOS (Toma et al. 2009). The expression for $\sin\theta'_B$, $\cos(2\phi'_B)$ and $\Delta\phi$ can be found in Toma et al. (2009) and Lan, Wu \& Dai (2016). The angle bracket denotes the average on the direction of the random magnetic field. $\chi_p$ is the PA of the polarized emission from a point-like region where the comoving LOS is fixed. $\chi_p=\phi+3\pi/2$ if $\bar{q}>0$, and $\chi_p=\phi$ if $\bar{q}<0$.

For the synchrotron emission in the ordered magnetic field model, we have (Toma et al. 2009),
\begin{equation}
F_{\nu}=\frac{1+z}{d_L^2}\frac{R^2A_0}{T_0}\int_0^{\theta_j+\theta_V} \mathcal{D}^2f(\nu')\sin\theta d\theta\int_{-\Delta\phi}^{\Delta\phi} (\sin\theta'_B)^{\alpha_1+1}d\phi ,
\end{equation}
\begin{equation}
Q_{\nu}=\frac{1+z}{d_L^2}\frac{R^2A_0}{T_0}\int_0^{\theta_j+\theta_V}\pi_0 \mathcal{D}^2f(\nu')\sin\theta d\theta\int_{-\Delta\phi}^{\Delta\phi} (\sin\theta'_B)^{\alpha_1+1}\cos(2\chi_p)d\phi ,
\end{equation}
\begin{equation}
U_{\nu}=\frac{1+z}{d_L^2}\frac{R^2A_0}{T_0}\int_0^{\theta_j+\theta_V}\pi_0 \mathcal{D}^2f(\nu')\sin\theta d\theta\int_{-\Delta\phi}^{\Delta\phi} (\sin\theta'_B)^{\alpha_1+1}\sin(2\chi_p)d\phi .
\end{equation}
The expressions of $\sin\theta'_B$ and $\chi_p$ for both aligned and toroidal magnetic fields can be found in Lan, Wu \& Dai (2016). $U_\nu$ is proved to be zero for the toroidal configuration (Toma et al. 2009; Lan, Wu \& Dai 2016), but can be non-zero for the aligned magnetic field case (Lan, Wu \& Dai 2016).

\subsection{AGN Case}

We assume that the emission region of a relativistic jet in an AGN is a thin shell which is similar to that in GRBs (i.e., the width of the emission region is much less than the radius of the jet). The central engine of AGNs is believed to be a SMBH plus an accretion disk. Such a system can launch a jet through the BZ mechanism, Blandford-Payne mechanism (BP; Blandford \& Payne 1982) or magnetic tower mechanism (Lynden-Bell 2003). The magnetic field configuration in the emission region of the jet powered by these mechanisms is very likely to be toroidal. However, considering that the jet might be accelerated by magnetic reconnection which in turn would disturb the ordered magnetic field, a random MFC is also expected. We therefore investigate the SOT and the SR models for AGNs.
For the SR model, we assume that the random magnetic field is confined in the shock plane. The flux and Stokes parameters of the emission in the SR model are expressed by
\begin{equation}
F_{\nu}=\frac{1+z}{4\pi d_L^2}\frac{\sqrt{3}e^3}{m_ec^2}B'\int_0^{\theta_j+\theta_V} d\theta \mathcal{D}^3\sin\theta<\sin\theta'_B>2\Delta\phi\int_{\gamma_{m}}^{\gamma_{max}} d\gamma_e N(\gamma_e)F(x),
\end{equation}
\begin{equation}
Q_{\nu}=\frac{1+z}{4\pi d_L^2}\frac{\sqrt{3}e^3}{m_ec^2}B'\int_0^{\theta_j+\theta_V} d\theta \mathcal{D}^3\sin\theta<\sin\theta'_B>\pi_p\int_{-\Delta\phi}^{\Delta\phi} d\phi \cos(2\chi_p)\int_{\gamma_{m}}^{\gamma_{max}} d\gamma_e  N(\gamma_e)F(x),
\end{equation}
where $e$ and $m_e$ are the charge and mass of the electron, $c$ is the speed of light, and $B'$ is the strength of the magnetic field. We denote $\bar{q}=-\pi_0\langle (\sin\theta'_B)^{1-m}\cos(2\phi'_B)\rangle$.
The local PD $\pi_p$ is expressed as $\pi_p=|\bar{q}|/\langle (\sin\theta'_B)^{1-m}\rangle$ with the spectral index $m$ defined by $F_\nu\propto\nu^m$. Here in AGN case, we take $\pi_0=0.6$.
The expression of $\sin\theta'_B$ and $\cos(2\phi'_B)$ can be found in Toma et al. (2009) and Lan, Wu \& Dai (2016). The local PA can be expressed as $\chi_p=\phi+3\pi/2$ if $\bar{q}>0$, and as $\chi_p=\phi$ if $\bar{q}<0$.
$N(\gamma_e)$ is the energy spectrum of the electrons with $\gamma_{m}$ and $\gamma_{max}$ the minimum and maximum Lorentz factors, respectively. $F(x)$ is the dimensionless spectrum of the synchrotron emission with $x=\nu'/\nu'_c$, where the critical frequency of the electron with Lorentz factor $\gamma_e$ is $\nu'_c=eB'\langle\sin\theta'_B\rangle\gamma_e^2/2\pi m_ec$.

The flux and Stokes parameters of the emission in the SOT model are expressed by
\begin{equation}
F_{\nu}=\frac{1+z}{2\pi d_L^2}\frac{\sqrt{3}e^3}{m_ec^2}B'\int_0^{\theta_j+\theta_V} d\theta \mathcal{D}^3\sin\theta\int_{0}^{\Delta\phi} d\phi\sin\theta'_B\int_{\gamma_{m}}^{\gamma_{max}} d\gamma_e N(\gamma_e)F(x),
\end{equation}
\begin{equation}
Q_{\nu}=\frac{1+z}{2\pi d_L^2}\frac{\sqrt{3}e^3}{m_ec^2}B'\pi_0\int_0^{\theta_j+\theta_V} d\theta \mathcal{D}^3\sin\theta\int_{0}^{\Delta\phi} d\phi \sin\theta'_B\cos(2\chi_p)\int_{\gamma_{m}}^{\gamma_{max}} d\gamma_e  N(\gamma_e)F(x) .
\end{equation}
The characteristic frequency of the electron with Lorentz factor $\gamma_e$ in the SOT model is $\nu'_c=eB'\sin\theta'_B\gamma_e^2/2\pi m_ec$. The expression of $\sin\theta'_B$ and $\chi_p$ for the toroidal magnetic field can be found in Toma et al. (2009) and Lan, Wu \& Dai (2016).

The low energy peaks in the energy spectra of the AGNs are commonly thought to be generated by the synchrotron emission. Here, we focus on the optical emission, which is at the vicinity of the low energy peak. Therefore, the synchrotron emission is the dominant mechanism. Considering that inverse Compton cooling is also important for an electron energy spectrum, we take the spectrum for the cooling electrons (Massaro et al. 2004, 2006)
\begin{eqnarray}
N(\gamma_e)=
\begin{cases}
K(\gamma_e/\gamma_{c})^{-s}, &\gamma_{m}\leq\gamma_e\leq\gamma_{c}  , \cr
K(\gamma_e/\gamma_{c})^{-s-r\log(\gamma_e/\gamma_{c})}, &\gamma_{c}<\gamma_e\leq\gamma_{max} ,
\end{cases}
\end{eqnarray}
where $\gamma_{c}=[2\pi m_ec\nu_{peak}(1+z)/\gamma eB']^{1/2}$ is the cutoff Lorentz factor. $\nu_{peak}$ is the position of the low energy peak in AGN energy spectrum. $K$, $s$ and $r$ are all constant.

\subsection{The effect of precession on the polarization properties}
The above Eqs. (4) - (12) are expressed in the coordinate system $xyz$ with $z$ along the LOS. The $x$-axis is set in the direction along the projection of the jet axis on the plane of sky. Thus the $xyz$ system is moving with the precessing jet. For comparison, we need to transform the Stokes parameters to a global coordinate system $XYz$ with $X$-axis along the direction of the projection of the precession axis on the plane of sky. The transformation of Stokes parameters between two coordinate systems is accomplished by the rotation matrix,
\begin{equation}\left(\begin{matrix}F_\nu^X\\ Q^X_\nu\\ U^X_\nu\end{matrix}\right)
=\left(\begin{matrix}1&0&0\\ 0&\cos2\alpha&\sin2\alpha\\ 0&-\sin2\alpha&\cos2\alpha\end{matrix}\right)
\left(\begin{matrix}F_\nu\\ Q_\nu\\ U_\nu\end{matrix}\right),
\end{equation}
where $F_\nu^X$, $Q^X_\nu$ and $U^X_\nu$ are the Stokes parameters in the coordinate system $XYz$. $\alpha$ is the angle between $x$-axis and $X$-axis. Then the PD and the PA of the emission from the precessing jet in coordinate system $XYz$ can be expressed as
\begin{equation}
\Pi^X=\frac{\sqrt{Q^{X2}_\nu+U^{X2}_\nu}}{F_\nu^X} ,
\end{equation}
\begin{equation}
\chi^X=\frac{1}{2}\arctan\left(\frac{U^X_\nu}{Q^X_\nu}\right) .
\end{equation}

A carton picture of the precessing jet is shown in Fig. 1. $\theta_{obs}$ is the angle between the precession axis and LOS. We set $\alpha=0$ when $t_{obs}=0$, and $t_{obs}$ is the observational time. We denote the intersection point of the precession axis on the plane of sky as $``O"$. The intersection point of the jet axis in the plane of sky is denoted as $``A"$. Because of the precession, the point $``A"$ is moving. Its initial position, i.e., when $t_{obs}=0$, is denoted as $``A_1"$. The angle between the two vectors, $\overrightarrow{OA_1}$ and $\overrightarrow{OA}$, is denoted as $\delta$. We then have $\delta=\Omega t_{obs}/(1+z)$, where $\Omega=2\pi(1+z)/T$ is the precessing angular velocity in the burst source frame. Thus, we have
\begin{equation}
\theta_V^2=\theta_i^2+\theta_{obs}^2-2\theta_i\theta_{obs}\cos\delta ,
\end{equation}
where $\theta_i$ is the angle between the precession axis and the jet axis. Then the $\alpha$ angle can be expressed as follows.
\begin{equation}
\cos\alpha=\frac{\theta_V^2+\theta_{obs}^2-\theta_i^2}{2\theta_V\theta_{obs}} ,
\end{equation}
where $\theta_V$ and $\theta_{obs}$ are both nonzero. We have $\theta_V=\theta_i$ when $\theta_{obs}=0$. In deriving Eqs. (17) and (18), the approximations $\theta_i \ll1$, $\theta_V\ll1$ and $\theta_{obs}\ll1$ are used.

Actually, the PD does not change during the rotation of the coordinate systems. Using Eqs. (14) and (15), we have
\begin{equation}
\Pi^X=\Pi^x .
\end{equation}
$\Pi^x$ is the PD in the coordinate system $xyz$. The PA after the rotation will be
\begin{equation}
\chi^X=\chi^x-\alpha+n\pi ,
\end{equation}
where $\chi^x$ is PA in the coordinate system $xyz$\footnote{Please note that $\chi^x$ is obtained by using the formula $\chi^x=1/2\arctan(U_\nu/Q_\nu)$ and also the signs of the Stokes parameters (Lan, Wu \& Dai 2018).}, and $n$ is an integer to eliminate the $180^\circ$ artificial jumps in PA curve.

\section{Numerical Results}
\subsection{GRB Case}
In the GRB case, two kinds of CCOs are involved, i.e., a black hole and a millisecond magnetar.
In the BH model, a jet is more likely powered by the BZ mechanism (Lei et al. 2017). The corresponding MFC in the jet is toroidal (Spruit et al. 2001). For a magnetar model, the MFC in a GRB jet is very likely to be aligned (Spruit et al. 2001). However, a random MFC generated by the shock or the magnetic reconnection is also possible. Therefore, in the GRB case, we consider three models: SR, SOT and SOA.

We take typical values for the parameters, $E_p=\gamma E'_p/(1+z)=300$\,keV, $\alpha_s=-0.2$, $\beta_s=1.2$, $\theta_i=3^\circ$ and $\theta_j=6^\circ$. The precessing period is assumed to be $T=70$ ms, which is inferred from Stone et al. (2013). The orientation of the aligned magnetic field (if there is) is assumed to be $\pi/6$. The observational frequency is taken to be $100\,$keV. The source is assumed to be located at redshift $z=1$. We present numerical results for the GRB light curves and polarization evolution for three models (i.e., SR, SOT and SOA) in Figs. 2-4, where $\theta_{obs}=4^\circ$ is used.

Fig. 2 shows the results for the SR model. Because the jet axis evolves far from the LOS before $T/2$ and then begins to approaches it after $T/2$ symmetrically due to precession, both the normalized spectral flux and the PD curve show the symmetrical distribution before and after $T/2$.
Generally speaking, the spectral flux and PD evolution show anti-correlation for the SR model. When the observational angle is small (i.e., $\theta_V<\theta_j$), the normalized spectral flux is $\sim1$ and PD is roughly 0 (Toma et al. 2009). While large PD is obtained only for off-axis observation (i.e., $\theta_V\sim\theta_j+1/\gamma$) for the SR model (Waxman 2003; Toma et al. 2009) and the spectral flux drops sharply when the LOS evolves to the outside of the jet cone. Although the PD value is non-negligible, the spectral flux is almost zero which leads to the PD detection very difficult around $T/2$.

PA of Fig. 2 evolves gradually and then changes abruptly by $\sim90^\circ$ before $T/2$. This abrupt $\sim90^\circ$ change of the PA happens when the majority of the observational cone (i.e., $1/\gamma$ cone) evolves from inside of the jet cone to the outside due to precession, which leads to the change of the sign of the Stokes parameter $Q_\nu$ in the coordinate system $xyz$ from negative to positive (Sari 1999). That is, in the coordinate system $xyz$, $\chi^x(t_{obs,1})=\pi/2$ while $\chi^x(t_{obs,2})=0$, where $t_{obs,1}$ and $t_{obs,2}$ denote the observational time just before and after the abrupt $\sim90^\circ$ PA change\footnote{It is predicted that this $\sim90^\circ$ abrupt change of the PA happen roughly at $\theta_V=\theta_j$, i.e., at $t_{obs}/T=\arccos((\theta_i^2+\theta_{obs}^2-\theta_j^2)/2\theta_i\theta_{obs})/2\pi\sim0.326$.}. Using Eq. (20) and noting that $-\pi/2<\alpha<0$, we have $\chi^X(t_{obs,1})=-\alpha(t_{obs,1})-\pi/2$ and $\chi^X(t_{obs,2})=-\alpha(t_{obs,2})$. Since the difference between $\alpha(t_{obs,1})$ and $\alpha(t_{obs,2})$ is quite small, the abrupt change of PA is roughly $90^\circ$. The abrupt $\sim90^\circ$ PA change after $T/2$ happens when the majority of the observational cone evolves from outside of the jet region to the inside. And the two abrupt $\sim90^\circ$ PA changes are symmetric in time before and after $T/2$ due to the symmetric observational geometry. We also notice that these two abrupt $\sim90^\circ$ changes of the PA happen when the PD changes from its decreasing phase to the rising phase, which is consistent with the conclusion of Lan, Wu \& Dai (2018).

The results for the SOT model are shown in Fig. 3. Same as that of Fig. 2, both the spectral flux and PD evolution show symmetric profile before and after $T/2$ due to precession. Different from that of the SR model, the spectral flux and PD curve are roughly positively correlated for the SOT model. Because when $\theta_V<\theta_j$, the observational cone is always within the jet region and the MFC in the observational cone is approximately aligned, the spectral flux and PD value are both high at the beginning. Then roughly when $\theta_V\sim\theta_j$ (i.e., at $t_{obs}/T=\arccos((\theta_i^2+\theta_{obs}^2-\theta_j^2)/2\theta_i\theta_{obs})/2\pi\sim0.326$), both the spectral flux and PD begin to decay, these are because part of the observational cone begins to be not covered by the jet cone and high latitude emission contribute significantly to the observed emission\footnote{PD of the high latitude emission for the SOT model is lower.}. In our calculation, $y_j\equiv(\gamma\theta_j)^2\sim100$. The evolution features of PD are consistent with that of Toma et al. (2009). The evolution of the PA is however gradual. In coordinate system $xyz$, the orientation of the ordered magnetic field in the observational cone is fixed which leads to the value of $\chi^x$ unchanged. Because of precession, $x-$axis is moving gradually relative to $X-$axis, so does the orientation of the ordered magnetic field in $XYz$ system. Since the change of the PA reflects the change of the direction of the ordered magnetic field in the $1/\gamma$ cone, PA ($\chi^X$) also evolves gradually during each period and the maximum value $\arcsin(\theta_i/\theta_{obs})\sim0.848$ rad (with $\theta_i<\theta_{obs}$) of the PA in $XYz$ system is reached at $t_{obs}/T=\arccos(\theta_i/\theta_{obs})/2\pi\sim0.115$ (i.e., at $d\alpha/dt_{obs}=0$). And the minimum PA value $-\arcsin(\theta_i/\theta_{obs})\sim-0.848$ rad is reached at $t_{obs}/T=1-\arccos(\theta_i/\theta_{obs})/2\pi\sim0.885$.

Fig. 4 presents the results for the SOA model. Both the spectral flux and the PD curve are symmetric before and after $T/2$ due to precession. Same as that in the SOT model, the spectral flux and PD curve are roughly positively correlated in the SOA model because of similar reason as that for the SOT model. But, the PA behaves differently. There is a PA bump during the valley of the spectral flux (see the lower panel of Fig. 4). The gradual increase phase of this PA bump begins at the vicinity of the time when the observational angle $\theta_V$ crosses $\theta_j+1/\gamma$.

Since the MFCs do not affect the spectral flux significantly, the light curves are almost the same in three models (i.e., SR, SOT and SOA). From Eq. (17), we know that the observational angle $\theta_V$ increases with $t_{obs}$ when $0<t_{obs}<T/2$ (as the jet axis goes away from the LOS) and then decreases during $T/2<t_{obs}<T$ (as the jet axis approaches the LOS). As a result, both the observed spectral flux and PD curve are symmetric before and after $T/2$. In the three models (especially in the SR and SOT models), the PA can change gradually with the precessing jet.

\subsection{TDE Case}
The central engines for TDEs are SMBHs with transient accretion disks. The detection of Sw J1644+57 at $z=0.3534$ suggested that a least some TDEs can launch a relativistic jet. Like AGNs, jets in TDEs might also be driven by the BZ mechanism, the BP mechanism or the magnetic tower mechanism. The MFCs in the TDE jets are also likely to be toroidal. On the other hand, the shock, turbulence and magnetic dissipation might happen during the jet propagation and disturb the ordered magnetic field. For the same reason, a random configuration is also possible. We therefore consider two models, i.e. SOT and SR.

We then numerically calculate the light curves and the polarization evolution of the TDE jet in two period of the precession. $E_p=\gamma E'_p/(1+z)=1.62$ keV, $\alpha_s=-0.33$, $\beta_s=1.12$, $\theta_i=5^\circ$, $\theta_j=10^\circ$, $T=2.7$ day and $z=0.3534$ are adopted in the calculations. The redshift, precessing period and the spectrum parameters are inferred by the observations of the Sw J1644+57 (Burrows et al. 2011). We study the emission at 2 keV.

We take observational angle $\theta_{obs}=8^\circ$ in Fig. 5. In our calculation, three bulk Lorentz factors are considered, i.e., $\gamma=10,\ 15,\ 20$. The observational cone is therefore roughly $1/\gamma\sim(2.9^\circ-5.7^\circ)$. The spectral flux and PD evolution are roughly anti-correlated for both SOT and SR models. With our setup, the observational angle increase before $T/2$ and then decrease after $T/2$ leading to a decreasing spectral flux before $T/2$ and an increasing spectral flux after $T/2$ for both SOT and SR models. For the SOT model, with the precession (i.e., with the increase of the observational angle $\theta_V$), the asymmetry of the observational geometry will increase and then decrease leading to an increase and then decrease of the observed PD before $T/2$. For the SR model, initially the observational cone is within the jet region and the corresponding PD is roughly 0. Then with the precession (i.e., with an increase $\theta_V$ before $T/2$ in our setup), the LOS will evolve to the outside of the jet region (leading to off-axis observation) and the resulting PD will rise before $T/2$. Because the spectral flux around $T/2$ is nearly zero, the PD around this time is difficult to be observed for both models.

Finally, we notice that the PA in the SOT and SR models can also change gradually with the precessing jet in the TDE case of Fig. 6. The evolution of the PA in TDE case can be analysed through the same way as that discussed in the GRB case because the PA evolution is also caused by the precession in the TDE case. For the SOT model, the maximum value $\arcsin(\theta_i/\theta_{obs})\sim0.675$ rad of the PA is reached at $t_{obs}/T=\arccos(\theta_i/\theta_{obs})/2\pi\sim0.143$ before $T/2$. For the SR model, the abrupt $\sim90^\circ$ change of the PA happens roughly at $t_{obs}/T=\arccos((\theta_i^2+\theta_{obs}^2-\theta_j^2)/2\theta_i\theta_{obs})/2\pi\sim0.272$. And also the abrupt $\sim90^\circ$ PA change happens when the PD evolves from its decline phase to rise phase.

\subsection{AGN Case}
The observations show that some AGNs have relativistic jets. Two possible MFCs in the emission region are toroidal and random, as described in Sec. 2.2. The SOT and SR models are considered in the AGN case. Parameters are $\theta_j=3.44^\circ$, $\theta_i=1.72^\circ$, $T=2$ yr and $B'=0.3$ G. The observational frequency is taken to be in the optical band with $\nu=4.365\times10^{14}$ Hz. The parameters for the electron energy spectrum are $\gamma_{min}=100$, $\gamma_{max}=10^6$, $\nu_{peak}=4\times10^{14}$ Hz, $s=2.3$ and $r=0.75$. The source is assumed to be located at redshift $z=1$. The local PD ($\pi_0$) is taken to be 0.6 for optical band.

Fig. 6 shows the light curves and polarization evolution in the AGN case with $\theta_{obs}=2.87^\circ$. The evolution of PA is also the result of jet precession and can be analyzed by the method used in GRB case. For the SOT model, the PA reaches its maximum value $\arcsin(\theta_i/\theta_{obs})\sim0.645$ rad when $t_{obs}/T=\arccos(\theta_i/\theta_{obs})/2\pi\sim0.147$. For the SR model, PA shows no abrupt $\sim90^\circ$ change for $\gamma=10$. When $\gamma=10$, the $1/\gamma\ (\sim5.7^\circ)$ cone is greater than the jet cone $\theta_j=3.44^\circ$. In one period, the majority of the $1/\gamma$ cone is always outside of the jet region. So there is no change of the sign of Stokes parameter $Q_\nu$ in $xyz$ frame and therefore no abrupt $\sim90^\circ$ PA change for $\gamma=10$. The abrupt $\sim90^\circ$ PA changes for $\gamma=20$ and 50 also occur when the PDs change from its decline phase to increase phase. These abrupt PA changes often happen when the $1/\gamma$ cone crosses the jet edge (i.e., when $\theta_V\sim\theta_j$).

\section{Conclusions and Discussion}
Several $\gamma-$ray polarimeters are now in operation, such as the International Gamma-Ray Astrophysics Laboratory/IBIS (INTEGRAL/IBIS; Winkler et al. 2003) and Hitomi (Aharonian et al. 2018). Furthermore, several X-ray polarimeteres are now in commission e.g., AstroSat (Singh et al. 2014), PoGo+ (Friis et al. 2018) and X-Calibur (Kislat et al. 2018). There will be a number of new X-ray polarimeters in near future, such as the Imaging X-ray Polarimetry Explorer (IXPE; Weisskopf et al. 2014), the enhanced X-ray Timing and Polarimetry (eXTP; Zhang et al. 2016) and the X-ray Polarization Probe (XPP). Many optical polarimeters are now in commission. There will be abundant polarization data in multi-bands for different astrophysical phenomenons. A dedicated theoretical study on the polarization involving different kinds of events is therefore highly demanded.

A precessing jet is naturally expected for central engines of GRBs, TDEs and AGNs, since an accretion disk (especially its outer part) is likely misaligned with the equatorial plane of the spinning CCO (Sarazin et al. 1980; Lu 1990; Lu \& Zhou 2005). The quasi-periodic variations in the light curves of GRBs, TDEs and AGNs are believed to be caused by the precession of jets. Investigating the polarization evolution of a precessing jet is thus of greater interest. This is the motivation of our paper. Adopting a precessing jet model, we performed detailed calculations on the light curves and polarization evolution for GRBs, TDEs and AGNs. Three types of polarization models, i.e., SR, SOT and SOA, are considered. Indeed, both the light curves and polarization properties exhibit a periodic nature.

The precession leads to a periodically changing observational angle (except that for $\theta_{obs}=0$). A direct result is the periodic nature of the observed flux and polarization. The symmetric features of both the light curve and the PD curve before and after $T/2$ are due to precession, i.e., they are geometric effect and have nothing to do with the polarization models, the jet bulk Lorentz factors and the kinds of astrophysical events. We found that the PA changes gradually even for SR and SOT models. This is unique for a precessing jet, since the PA of a non-precessing top-hat jet can only change abruptly by $90^\circ$ for these two models. Therefore, future high quality polarization data can be used to differentiate the precessing jet and non-precessing jet, as well as the MFCs.

\acknowledgements
This work is supported by the National Key Research and Development Program of China (grant no. 2017YFA0402600) and the National Natural Science Foundation of China (grant no. 11573014, 11673068, 11725314, 11433009, 11773010, and 11833003). X.F.W. is also partially supported by the Key Research Program of Frontier Sciences (QYZDB-SSW-SYS005), the Strategic Priority Research Program ``Multi-waveband gravitational wave Universe'' (grant No. XDB23040000) of the Chinese Academy of Sciences and the ``333 Project" of Jiangsu province. M.X.L is supported by the Natural Science Foundation of Jiangsu Province (grant No. BK20171109).

\begin{figure}
\centering
\includegraphics[angle=0,scale=0.6]{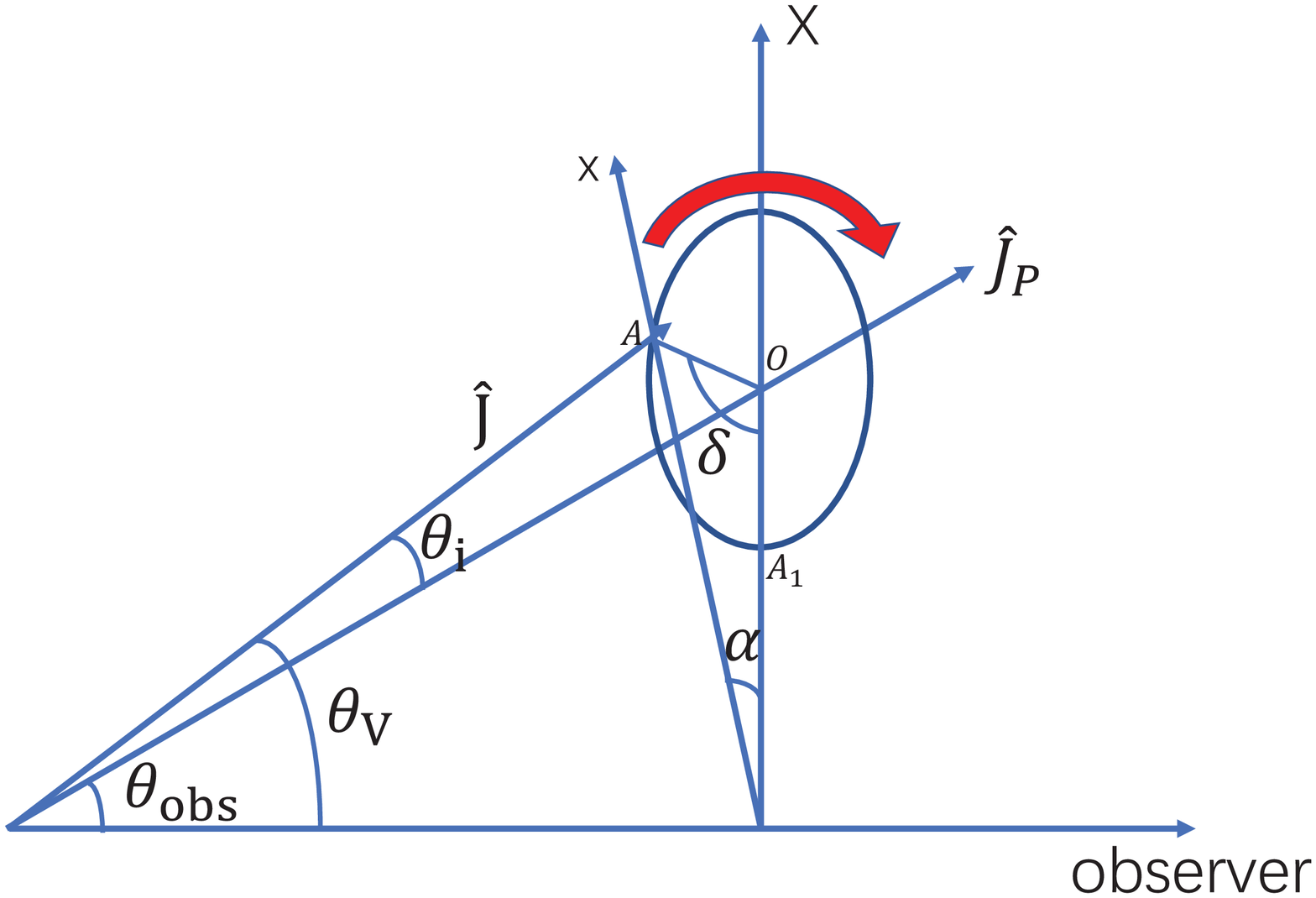}
\caption{A sketch figure of the precessing jet and the coordinate systems used in our calculations. $\hat{J}_p$ is the precession axis, $\hat{J}$ is the jet axis, the red arrow points toward the precessing direction of the jet, $\theta_{\rm obs}$ is the angle between the precession axis and the LOS, $\theta_{\rm V}$ is the observational angle, $\alpha$ is the angle between the $x-$axis and the $X-$axis. \label{fig1}}
\end{figure}

\begin{figure}
\includegraphics[angle=0,scale=0.8]{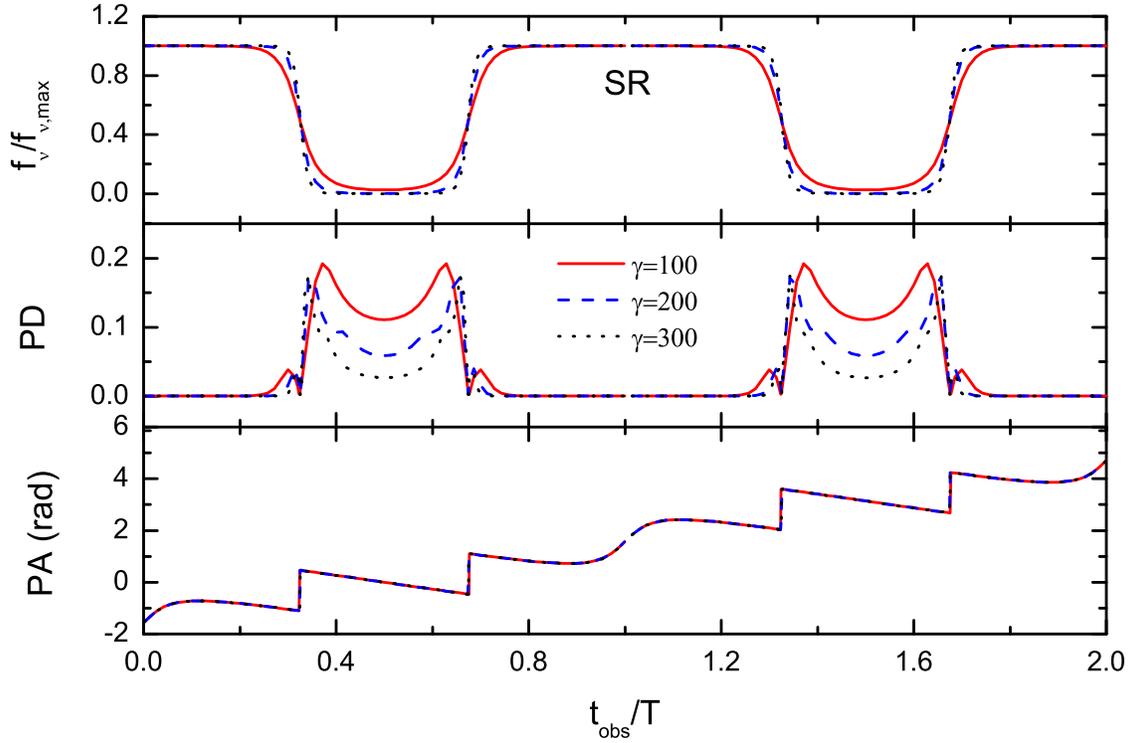}
\caption{Gamma-ray light curves (upper panel) and the evolution of PD (middle panel) and PA (lower panel) of a precessing GRB jet for the synchrotron emission in a random magnetic field (SR) model and $\theta_{obs}=4^\circ$, and for different Lorentz factor $\gamma=100$ (solid lines), 200 (dashed lines) and 300 (dotted lines). $E_p=300$ keV, $\alpha_s=-0.2$, $\beta_s=1.2$, $\theta_i=3^\circ$ and $\theta_j=6^\circ$ are adopted in the calculation. In the upper panel, the flux are normalized. \label{fig2}}
\end{figure}

\begin{figure}
\includegraphics[angle=0,scale=0.6]{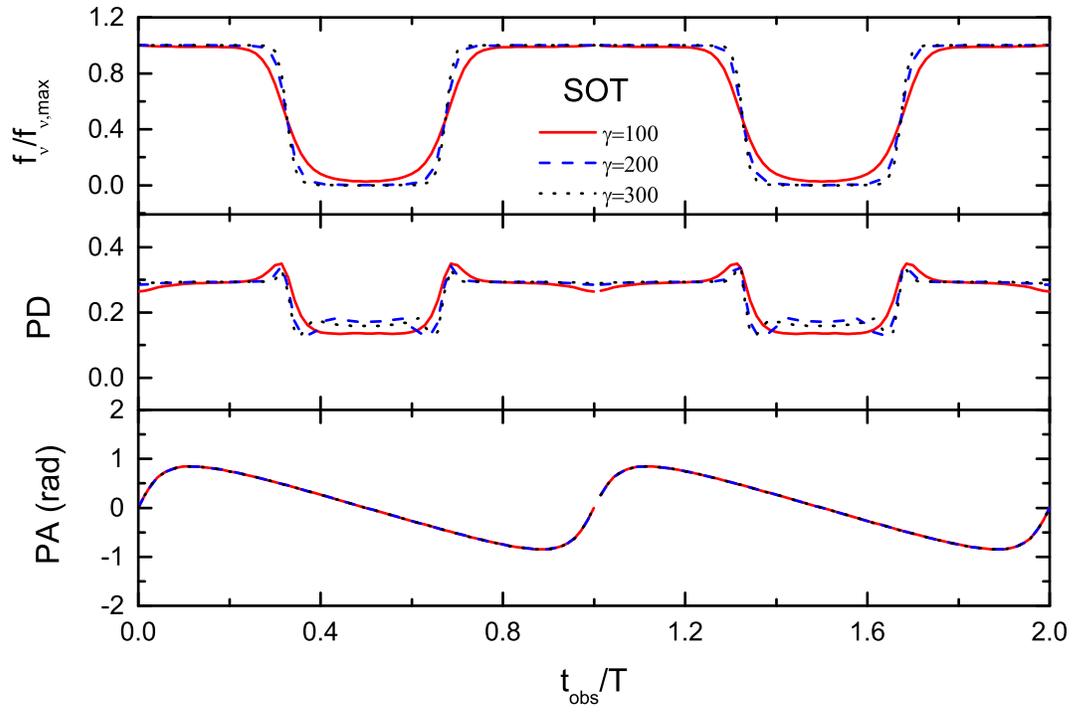}
\caption{Same as Fig. 2, but for the synchrotron emission in an ordered toroidal magnetic field (SOT) model. \label{fig3}}
\end{figure}

\begin{figure}
\includegraphics[angle=0,scale=0.8]{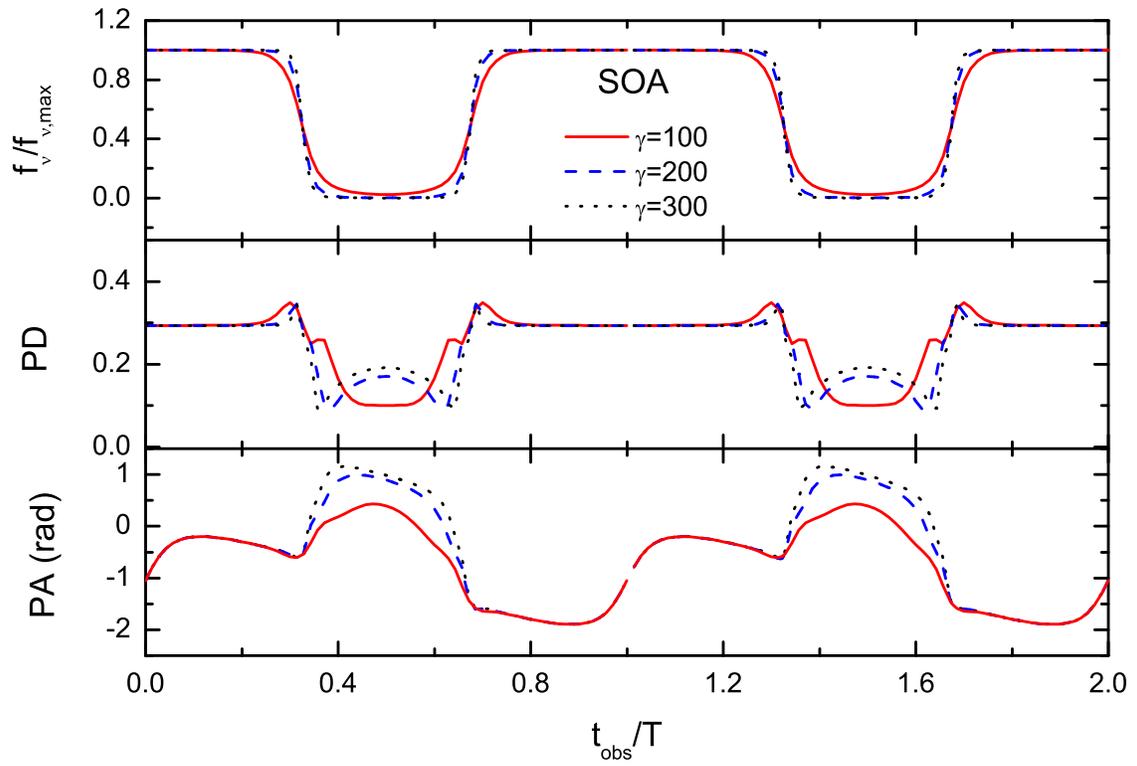}
\caption{Same as Fig. 2, but for the synchrotron emission in an ordered aligned magnetic field (SOA) model. \label{fig4}}
\end{figure}

\begin{figure}
\includegraphics[angle=0,scale=0.8]{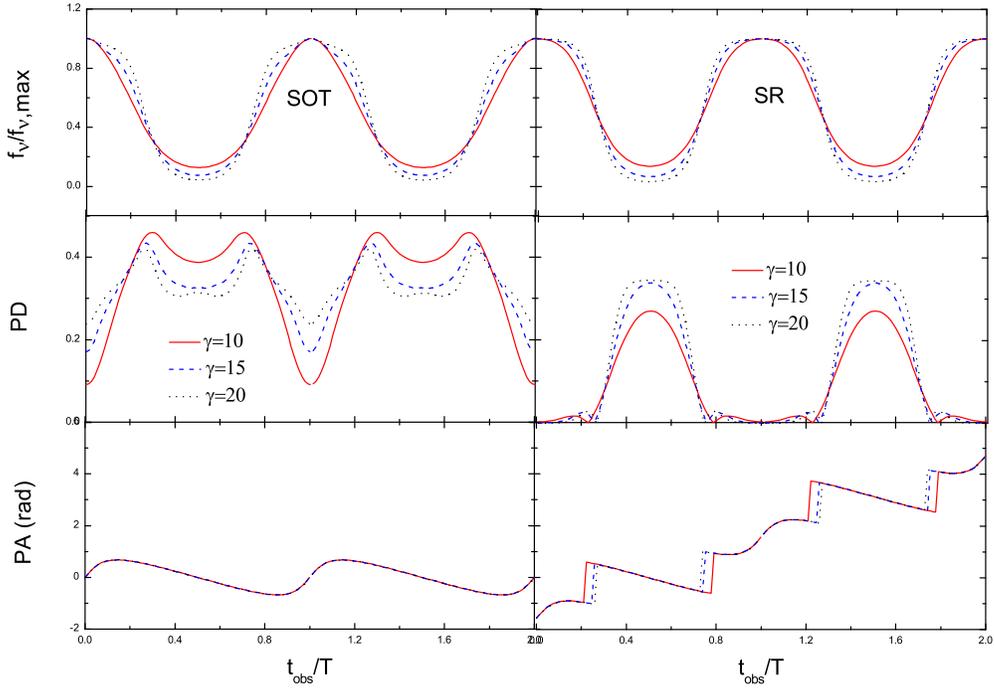}
\caption{X-ray light curves (upper panel) and the evolution of PD (middle panel) and PA (lower panel) of a precessing TDE jet for the SOT (left) and SR (right) model and $\theta_{obs}=8^\circ$, and for different Lorentz factor $\gamma=10$ (solid lines), 15 (dashed lines) and 20 (dotted lines). $E_{\rm p}=1.62$ keV, $\alpha_s=-0.33$, $\beta_s=1.12$, $\theta_i=5^\circ$, $\theta_j=10^\circ$, $T=2.7$ day and $z=0.3534$ are adopted in the calculations.
\label{fig5}}
\end{figure}

\begin{figure}
\includegraphics[angle=0,scale=0.8]{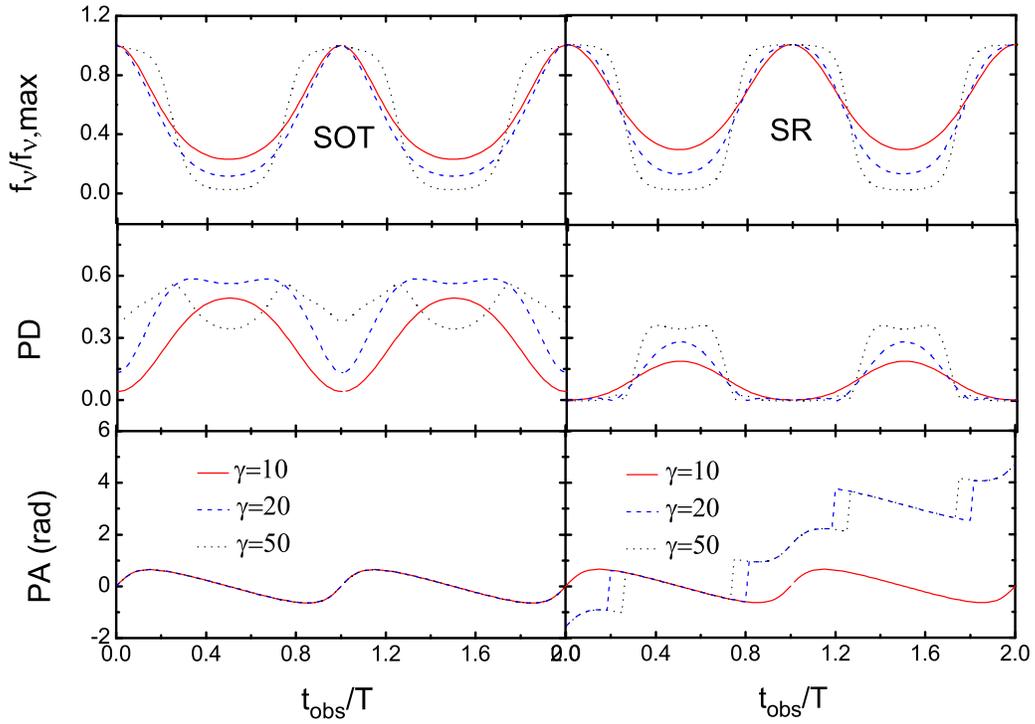}
\caption{Optical light curves (upper panel) and the evolution of PD (middle panel) and PA (lower panel) of a precessing AGN jet for the SOT (left) and SR (right) model and $\theta_{obs}=2.87^\circ$, and for different Lorentz factor $\gamma=10$ (solid lines), 15 (dashed lines) and 20 (dotted lines). $\gamma_{min}=100$, $\gamma_{max}=10^6$, $\nu_{peak}=4\times10^{14}$ Hz, $s=2.3$, $r=0.75$ and $z=1$ are used.
\label{fig7}}
\end{figure}

\end{document}